\documentclass[12pt,fleqn]{article}

\usepackage{amsmath,amssymb,amsfonts,graphicx,epsfig,cite}
\def\bea{\begin{eqnarray}}
\def\eea{\end{eqnarray}}
\def\be{\begin{equation}}
\def\ee{\end{equation}}
\def\ba{\begin{array}}
\def\ea{\end{array}}
\def\nn{\nonumber}
\def \lsim{\mathrel{\vcenter
{\hbox{$<$}\nointerlineskip\hbox{$\sim$}}}}
\def \gsim{\mathrel{\vcenter
{\hbox{$>$}\nointerlineskip\hbox{$\sim$}}}}

\def\a{&\hspace{-7pt}}
\def\b{&\hspace{-10pt}}

\begin{document}
\pagestyle{plain}

\renewcommand{\theequation}{\arabic{section}.\arabic{equation}}
\setcounter{page}{1}

\begin{titlepage} 
\begin{center}

$\;$

\vskip 0.5 cm

\begin{center}
{\LARGE{ \bf Simple metastable de Sitter vacua \\[2mm] in N=2 gauged supergravity}}
\end{center}

\vskip 1cm

{\large 
Francesca Catino${}^{a,b}$, Claudio A. Scrucca${}^c$ and Paul Smyth${}^c$
}

\vskip 0.5cm

$^{a}${\it II. Institut f\"ur Theoretische Physik, Universit\"at Hamburg, \\
D-22761 Hamburg, Germany}\\[2mm]
$^{b}${\it Zentrum f\"ur Mathematische Physik, Universit\"at Hamburg,\\
Bundesstrasse 55, D-20146 Hamburg, Germany}\\[2mm]
${}^c${\it Institut de Th\'eorie des Ph\'enom\`enes Physiques, EPFL, \\
\mbox{CH-1015 Lausanne, Switzerland}\\
}

\vskip 1.5cm

\begin{abstract}

We construct a simple class of N=2 gauged supergravity theories that admit 
metastable de Sitter vacua, generalizing the recent work done in the context 
of rigid supersymmetry. The setup involves one hypermultiplet and one vector 
multiplet spanning suitably curved quaternionic-K\"ahler and special-K\"ahler geometries,
with an Abelian gauging based on a single triholomorphic isometry, but neither 
Fayet-Iliopoulos terms nor non-Abelian gauge symmetries. We construct the most 
general model of this type and show that in such a situation the possibility of achieving 
metastable supersymmetry breaking vacua crucially depends on the value of the 
cosmological constant $V$ relative to the gravitino mass squared $m^2_{3/2}$ in Planck units. 
In particular, focusing on de Sitter vacua with positive $V$, we show that metastability 
is only possible when $V \gsim 2.17\, m^2_{3/2}$. We also derive an upper bound on 
the lightest scalar mass in this kind of model relative to the gravitino mass $m_{3/2}$ 
as a function of the cosmological constant $V$, and discuss its physical implications. 

\end{abstract}

\bigskip

\end{center}
\end{titlepage}

\newpage

\section{Introduction}
\setcounter{equation}{0}

Supersymmetry, if it is realized in Nature, must be spontaneously broken 
on a ground state that is at least metastable, in such a way to comply with 
the negative experimental searches performed so far. In the context of 
supergravity theories, this has motivated several systematic studies of 
the circumstances under which metastable non-supersymmetric vacua 
may arise. Finding such vacua turns out to be surprisingly difficult, even 
when allowing the freedom of adjusting all the parameters of the theory, because
the scalar potential has a restricted functional form in supersymmetric 
theories. This difficulty moreover increases with the number of supersymmetries.

One general strategy that can be pursued to uncover possible obstructions
against metastable supersymmetry breaking in generic supergravity theories consists 
of studying the general structure of the masses of the sGoldstini, which represent 
an endemic danger of instability. 
For N=1 theories, this study has been performed 
exhaustively in \cite{GRS1,GRS2,GRS3} (see also \cite{DD,CGRGLPS1,CGRGLPS2}). 
The outcome is that metastable de Sitter vacua may exist only if the scalar manifold 
has a suitable curvature. For N=2 theories, the same study has only been carried out in 
certain special classes of theories. In particular, it has been shown that metastable de 
Sitter vacua are ruled out in theories with only hypermultiplets \cite{GRLS} or only Abelian 
vector multiplets \cite{MANY}. On the other hand, for the more general classes of theories 
with charged hypermultiplets or non-Abelian vector multiplets a few examples of metastable 
de Sitter vacua are known \cite{FTV,O}, but no constraint on the existence of such 
vacua has been worked out so far. For N=4 and N=8 theories, a similar study has been 
done in \cite{BR,BLR}, where it was proven that metastable de Sitter vacua could 
only arise in a very constrained region of parameter space. No genuine examples 
of such vacua are known in this context, but their existence has not yet been excluded 
(examples of unstable de Sitter vacua \cite{DWP,DWPT,HW,KLPS,DI2} and marginally 
stable Minkowski vacua possessing flat directions \cite{DI1} are known). 
Finally, the same kind of study has also been applied to 
theories defined through a truncation that reduces the amount of supersymmetry, where an unstable 
de Sitter vacuum of the original theory may lead to a metastable de Sitter vacuum in the truncated theory. 
The simplest case of N=2 to N=1 truncations with only scalar multiplets has been studied in \cite{CSS}, 
and several non-trivial examples in the context of N=8 to N=4 and N=4 to N=2 were described in 
\cite{RR}. 

The same issue of vacuum metastability when supersymmetry is spontaneously broken 
already arises in the simpler context of theories with global supersymmetry. 
One may then also study the problem in this simpler context, without losing any of its essential 
features, at least for N=1 and N=2 theories where, contrary to N=4 and N=8 theories, one 
can have a generic scalar geometry even in the rigid limit. Moreover,  for simplicity one may discard 
constant Fayet-Iliopoulos terms, since these are essentially an accidental feature of rigid 
supersymmetry and are only compatible with gravity under quite restrictive circumstances. 
This rigid version of the metastability problem was studied in some detail in \cite{JS,BS},  again using the strategy of
looking at the masses of the sGoldstini. The rigid limit of all the results known from the supergravity 
analyses for generic N=1 and special N=2 theories were recovered in a simpler and more transparent 
way. Furthermore, it was argued that in N=2 theories with non-Abelian vector multiplets 
there is no obstruction against metastability from the sGoldstini. On the other hand, no constraint 
has been derived so far with this approach for the general case of N=2 theories with charged 
hypermultiplets. 

In the context of rigid supersymmetry, one can also use other types 
of analyses that better exploit the control that one has over the off-shell theory. For instance, assuming 
that metastable supersymmetry breaking is possible and that the only massless states are
the Goldstini, one can try to explicitly construct their low-energy effective theory in which 
supersymmetry is realized non-linearly. Any obstruction showing up 
in this attempt can then be interpreted as signaling the impossibility of realizing the assumed
metastable supersymmetry breaking vacuum. Proceeding along these lines, it was shown 
in \cite{AB} that N=2 theories possessing an $SU(2)_R$ global symmetry cannot admit 
a genuinely metastable supersymmetry breaking vacuum, at least under the further, more 
technical assumption that they admit a well-defined supercurrent superfield satisfying a 
conservation law that involves at most a superconformal linear anomaly multiplet. This 
result strongly suggests that a crucial requirement for a generic N=2 theory to admit a viable 
metastable non-supersymmetric vacuum is that it should not possess any global  $SU(2)_R$ symmetry. 
Remarkably, the presence of such an $SU(2)_R$ symmetry 
rests on radically different features in the hypermultiplet and vector multiplet sectors, 
and this introduces an important distinction between them. In the vector multiplet sector,
an $SU(2)_R$ symmetry automatically emerges whenever constant Fayet-Iliopoulos terms are absent, 
even if the scalar manifold is arbitrarily curved. In the hypermultiplet sector,
on the other hand, no $SU(2)_R$ symmetry can arise if one considers a sufficiently 
generic curved scalar manifold. As a consequence, the simplest 
candidates for N=2 theories admitting metastable non-supersymmetric vacua are those 
involving charged hypermultiplets. Following this expectation, it has been shown in \cite{LSS} 
that the simplest class of such theories based on just one hypermultiplet and one Abelian 
vector multiplet with suitably curved scalar manifolds does indeed admit metastable 
non-supersymmetric vacua.

The aim of this work is to construct a simple, minimal class of N=2 supergravity theories
that admit metastable de Sitter vacua, without involving either constant Fayet-Iliopoulos 
terms or non-Abelian gaugings, by generalizing the construction presented in \cite{LSS} 
from rigid to local supersymmetry. The setup involves one hypermultiplet and one vector 
multiplet spanning suitably curved quaternionic-K\"ahler and special-K\"ahler geometries, 
with an Abelian gauging based on a single triholomorphic isometry. Our aim is to construct 
the most general model of this type and study the circumstances under which this admits a 
metastable supersymmetry breaking vacuum. We shall see how this can be achieved by
fixing a point in the scalar manifold and then tuning the geometry in the neighborhood
of that point such that it corresponds to a metastable de Sitter vacuum.
As already argued in \cite{LSS}, this possibility 
could a priori depend on the value of the cosmological constant $V$ relative to the gravitino 
mass squared $m^2_{3/2}$ in Planck units, and the relevant dimensionless parameter is therefore
expected to be given by the following expression, which we restrict for simplicity to be positive:
\be
\epsilon = \frac {V}{m^2_{3/2}} \,.
\label{epsilon}
\ee
More precisely, when $\epsilon \gg 1$ the influence of gravitational effects on scalar masses 
is negligible. Therefore, by virtue of the results derived in \cite{LSS} for the rigid limit, it should
be possible to achieve a viable metastable vacuum by adjusting the form of the scalar manifolds.
On the other hand, when $\epsilon \ll1$ the influence of gravitational effects on scalar masses 
is a priori significant and the possibility to achieve a viable metastable vacuum by adjusting the 
form of the scalar manifolds must be reexamined within supergravity.

The rest of this paper is structured as follows. In section 2 we briefly review the main features 
of  N=2 gauged supergravity theories that will be relevant for our purposes. In particular, we describe
useful explicit parametrizations of the most general quaternionic-K\"ahler and special-K\"ahler 
manifolds of minimal dimensions four and two in terms of Toda and Laplace potentials, 
respectively, which are the main building blocks of our model. In sections 3 and 4 we then review the known properties 
of the models with only one hypermultiplet and only one vector multiplet that can be constructed 
with these spaces, and recall specifically the structure of their scalar mass matrices and the sum 
rules forbidding the existence of metastable vacua. In section 5 we then explicitly construct 
the class of models involving one hypermultiplet and one vector multiplet that we are interested in. 
Next we compute the structure of the scalar mass matrix and study the constraints that can be put 
on its eigenvalues. Finally, we derive a sharp upper bound on the smallest scalar mass 
as a function of the parameter $\epsilon$, and deduce from this the range of values for $\epsilon$ 
for which metastable vacua are allowed. In section 6, we briefly discuss the rigid 
limit of our supergravity analysis and show how it matches the rigid supersymmetry analysis 
of \cite{LSS}. In section 7, we then present a class of explicit examples of models admitting 
metastable de Sitter vacua and compute for these the full spectrum of scalar masses. 
In section 8 we present our conclusions.

\section{N=2 gauged supergravity}
\setcounter{equation}{0}

Let us briefly review the general structure of N=2 gauged supergravity theories, 
restricting to Abelian symmetries and using Planck units. In general, there can be $n_H$ hypermultiplets 
and $n_V$ vector multiplets. The $4\, n_H$ real scalars $q^u$ from the hypermultiplets 
span a quaternionic-K\"ahler manifold with metric $g_{uv}$ while the $n_V$ complex scalars 
$z^i$ from the vector multiplets span a special-K\"ahler manifold with metric $g_{i \bar \jmath}$. 
The $n_V$ vectors $A_\mu^a$ from the vector multiplets and the graviphoton $A_\mu^0$, denoted 
altogether by $A_\mu^A$, have kinetic metric $\gamma_{AB} = - {\rm Im}\, {\cal N}_{AB}$ and topological 
angles $\theta_{AB} = {\rm Re}\, {\cal N}_{AB}$ in terms of the so-called period matrix ${\cal N}_{AB}$ 
associated with the special-K\"ahler manifold. They can be used to gauge a maximum of $n_V+1$ of isometries, 
which are described by triholomorphic Killing vectors $k_A^u$ on the quaternionic-K\"ahler 
manifold. The scalar and vector kinetic energy is given by \cite{AGF1,AGF2,BW,dWvP,dWLvP,HKLR,S,DFF}:
\bea
T = - \text{\small $\frac 14$} \gamma_{\hspace{-1pt}AB} F^A_{\hspace{-1pt}\mu \nu} F^{B\mu \nu} \!
+ \text{\small $\frac 14$} \theta_{\hspace{-1pt}AB} F^A_{\hspace{-1pt}\mu \nu} \tilde F^{B \mu \nu} \!
- \text{\small $\frac 12$} \hspace{1pt} g_{uv}\hspace{1pt} D_\mu q^u D^\mu q^v \!
- g_{i \bar \jmath}\, \partial_\mu z^i \partial^\mu \bar z^{\bar \jmath} \,.
\label{T}
\eea
In this expression $F^A_{\mu \nu} = \partial_\mu A^A_\nu - \partial_\nu A^A_\mu$, 
$\tilde F^A_{\mu \nu} = \frac 12\, \epsilon_{\mu \nu \rho \sigma} F^{A \rho \sigma}$ and
$D_\mu q^u = \partial_\mu q^u + k^u_A A^A_\mu$. The scalar potential is instead 
given by
\bea
V = 2\, g_{uv} k_A^u k_B^v L^A \bar L^B + g^{i \bar \jmath} f_i^A \bar f_{\bar \jmath}^B P_A^x P_B^x 
- 3 P_A^x P_B^x L^A \bar L^B \,.
\label{V}
\eea
Here $L^A$ denotes the covariantly holomorphic symplectic section of the special-K\"ahler manifold
and $f_i^A = \nabla_i L^A$, while $P_A^x$ denote the three Killing potentials admitted by 
each of the triholomorphic Killing vector $k_A^u$. Finally, let us also recall that the gravitino mass
is given by
\bea
m^2_{3/2} = P_A^x P_B^x L^A \bar L^B \,.
\eea

In the following, we will concentrate on the simplest cases where $n_H$ is $0$ or $1$ and 
$n_V$ is $0$ or $1$. To construct the most general theories of this type, we will need to consider 
the most general quaternionic-K\"ahler manifold of dimension four with at least one triholomorphic
isometry and the most general special-K\"ahler manifold of dimension two. Fortunately, there 
exist general local parametrizations for these two kinds of manifolds, in terms of potentials 
$h$ and $l$ of three and two real variables satisfying the Toda and the Laplace equations, respectively. 
It is then possible to construct general theories based on generic choices for these two functions. 

\subsection{Quaternionic-K\"ahler manifold}

In the hypermultiplet sector, we will consider a generic four-dimensional quaternionic-K\"ahler space 
admitting at least one triholomorphic isometry. With the canonical normalization adopted in (\ref{T})
for the scalar kinetic term (corresponding to $\lambda = - \frac 12$ in the notation of \cite{ABCDFFM}), 
the scalar curvature must be $R = -12$. It turns out that the line element of such a manifold can always 
be locally brought into the following Przanowski-Tod form \cite{P,T}, using four real coordinates $q^u = \rho,\varphi,\chi,\tau$:
\be
d s^2 = g_{uv} \, dq^u dq^v = \frac 1{2 \rho^2} \Big(f d\rho^2 + f e^h (d\varphi^2 + d\chi^2) + f^{\text{--}1} (d\tau + \Theta)^2 \Big) \,.
\ee
This depends on a single function $h$ of the three variables $q^{\hat u} = \rho, \varphi, \chi$, which must satisfy 
the three-dimensional Toda equation:
\be\label{toda-h}
(e^h)_{\rho\rho} + h_{\varphi\varphi} + h_{\chi\chi} = 0 \,.
\ee
The function $f$ is then related to the function $h$ by
\be
f = 2 - \rho\hspace{1pt} h_\rho \,.
\ee
The $1$-form $\Theta$ is instead determined, modulo an irrelevant exact form, by the following equation, whose 
integrability is guaranteed by the Toda equation:
\be
d \Theta = \big(f_\varphi \, d\chi - f_\chi \, d\varphi\big) \wedge d\rho + (f e^h)_\rho \, d\varphi \wedge d\chi \,.
\ee
Further details about the geometric properties of this space can be found for example in \cite{DSTV,CSS}.
The $SU(2)$ connection $\omega^x$ is found to be:
\bea
\b\b \omega^1 = - \rho^{\text{--}1} e^{h/2} d\chi \,,\\[1mm]
\b\b \omega^2 = - \rho^{\text{--}1} e^{h/2} d\varphi \,,\\ 
\b\b \omega^3 = - \text{\small $\frac 12$} \rho^{\text{--}1} (d\tau + \Theta) 
+ \text{\small $\frac 12$} h_\chi d\varphi - \text{\small $\frac 12$} h_\varphi d\chi \,.
\eea
The three hyper-K\"ahler forms $J^x$, which satisfy $\nabla J^x =0$ as a consequence of the equation defining $\Theta$ 
and are thus only covariantly closed, are given by:
\bea
\b\b J^1 = \text{\small $\frac 12$} \rho^{\text{--}2} e^{h/2}d\varphi \wedge (d\tau + \Theta) 
- \text{\small $\frac 12$} \rho^{\text{--}2} f e^{h/2} d\rho \wedge d\chi \,, \\
\b\b J^2 = - \text{\small $\frac 12$} \rho^{\text{--}2} e^{h/2} d\chi \wedge (d\tau + \Theta) 
- \text{\small $\frac 12$} \rho^{\text{--}2} f e^{h/2} d\rho \wedge d\varphi \,, \\
\b\b J^3 = \text{\small $\frac 12$} \rho^{\text{--}2} d\rho \wedge (d\tau + \Theta) 
+ \text{\small $\frac 12$} \rho^{\text{--}2} f e^{h} d\varphi \wedge d\chi \,.
\eea 
Finally, the manifest isometry amounts to a shift in the variable $t$ and the corresponding Killing vector 
takes the following expression, involving an arbitrary parameter $\xi$ of dimension one:
\be
k = \xi\, \partial_\tau \,.
\ee

In the above parametrization in terms of the four real coordinates $q^u = \rho, \varphi,\chi,\tau$, the components
of the metric are:
\bea
g_{uv} = \frac 1{2\hspace{1pt} \rho^2} \left(\begin{matrix}
f \!+\! f^{\text{--}1} \Theta_\rho^2 & f^{\text{--}1} \Theta_\rho \Theta_\varphi & f^{\text{--}1} \Theta_\rho \Theta_\chi & f^{\text{--}1} \Theta_\rho \\
f^{\text{--}1} \Theta_\varphi \Theta_\rho  & f e^h \!+\! f^{\text{--}1} \Theta_\varphi^2 &  f^{\text{--}1} \Theta_\varphi \Theta_\chi & f^{\text{--}1} \Theta_\varphi \\
f^{\text{--}1} \Theta_\chi \Theta_\rho & f^{\text{--}1} \Theta_\chi \Theta_\varphi & fe^h \!+\! f^{\text{--}1} \Theta_\chi^2 & f^{\text{--}1} \Theta_\chi \\
f^{\text{--}1} \Theta_\rho & f^{\text{--}1} \Theta_\varphi & f^{\text{--}1} \Theta_\chi & f^{\text{--}1} \\
\end{matrix}\right) \,.
\eea
Positivity of this metric requires $f > 0$. A simple choice for the vielbein $e_u{}^p$ that can be used to locally trivialize this metric
as $g_{uv} = e_u{}^p \delta_{pq}\, (e^T)^q{}_v$ is given by:
\bea
\b\b e_u{}^p = \frac 1{\text{\small $\sqrt{2}$}\, \rho}\left(\begin{matrix}
f^{1/2} & 0 & 0 & f^{\text{--}1/2} \Theta_\rho \\
0 & f^{1/2} e^{h/2}& 0 & f^{\text{--}1/2} \Theta_\varphi \\
0 & 0 & f^{1/2} e^{h/2}& f^{\text{--}1/2} \Theta_\chi \\
0 & 0 & 0 & f^{\text{--}1/2}
\end{matrix}\right) \,.
\eea
The components of the three hyper-K\"ahler forms are also easily worked out and seen to satisfy
the algebra $(J^x)^u{}_w (J^y)^w{}_v = - \delta^{xy} \delta^u_v + \epsilon^{xyz} (J^z)^u{}_v$. One 
can then easily verify that the Killing vector with components $k^u$ is triholomorphic and compute 
the corresponding Killing potentials $P^x$ from their defining relation $\nabla_u P^x = - (J^x)_{uv} k^v$. 
The result is:
\bea
k^u = \left(\begin{matrix}
0 \\ 0 \\ 0 \\ \xi
\end{matrix}\right)\,,\;\; |\vec P| = \text{\small $\frac 12$}\, \xi \, \rho^{\text{--}1} \,.
\label{KillingPT}
\eea

It will be useful for the analysis of the forthcoming sections to define the following three-dimensional matrix of dimensionless 
parameters associated to the second derivatives of the function $f$ with respect to its variables $q^{\hat u} = \rho, \varphi, \chi$:
\bea
\alpha_{\hat u \hat v} = \text{\small $\frac 12$} (e^{\text{--}1})_{\hat u}{}^{\hat p} f_{\hat p \hat q} (e^{\text{--}1T})^{\hat q}{}_{\hat v} \,. 
\label{a}
\eea
A simple computation shows that the entries of this matrix are given by
\bea
\b\b \alpha_{\rho \rho} = f^{\text{--}1} \hspace{-1pt} \rho^2 f_{\rho\rho} \,,\;\; \\
\b\b \alpha_{\rho \varphi} = f^{\text{--}1} \hspace{-1pt} \rho^2 f_{\rho \varphi} e^{\text{--}h/2} \,,\;\; 
\alpha_{\rho \chi} = f^{\text{--}1} \hspace{-1pt} \rho^2 f_{\rho \chi} e^{\text{--}h/2} \,, \\
\b\b \alpha_{\varphi \varphi} = f^{\text{--}1} \hspace{-1pt} \rho^2 f_{\varphi\varphi} e^{\text{--}h} \,,\;\; 
\alpha_{\chi\chi} = f^{\text{--}1} \hspace{-1pt} \rho^2 f_{\chi \chi} e^{\text{--}h} \,,\;\; 
\alpha_{\varphi \chi} = f^{\text{--}1} \hspace{-1pt} \rho^2 f_{\varphi\chi} e^{\text{--}h} \,.
\eea
The definitions and properties of the functions $h$ and $f$ imply an important constraint on the second derivatives of $f$, 
and thus on the parameter $\alpha_{\hat u \hat v}$. More precisely, using the relation between $f$ and $h$ and the Toda 
equation (\ref{toda-h}) satisfied by $h$, one easily shows that
$f_{\rho \rho} + e^{\text{--}h} f_{\varphi \varphi} + e^{\text{--}h} f_{\chi\chi} = - \rho^{\text{--}2} f(f-1)(f-2) + \rho^{\text{--}1} (3f-4) f_\rho$.
As a consequence of this property, it follows that the trace of the matrix $\alpha_{\hat u \hat v}$ involving the second derivatives of $f$ 
is completely fixed in terms of the first derivatives of $f$ and the function $f$ itself, and one finds:
\bea
\delta^{\hat u \hat v} \alpha_{\hat u \hat v}  = - \big(f-1\big)\big(f-2\big) + \big(3 f - 4\big) f^{\text{--}1}\! \rho f_\rho \,.
\label{constrainta}
\eea

\subsection{Special-K\"ahler manifold}

In the vector multiplet sector, we will consider a general two-dimensional special-K\"ahler manifold of the local type.
This can be locally described by a metric of the following form, using the coordinates $q^\alpha = z,\bar z$:
\bea
ds^2 = g_{\alpha \bar \beta} \, dq^\alpha d q^{\bar \beta} = 2\, g_{z \bar z} \, dz \, d \bar z = 2\,l\, |d z|^2 \,.
\eea
The line element depends on a single real function $l$ of $z$ and $\bar z$. This function is restricted by 
the fact  that it can be expressed in terms of a holomorphic prepotential. 
The resulting restriction on the curvature of the space can be taken into account efficiently in terms of the symplectic section
$L^A(z,\bar z)$, where $A=0,1$. In the local case, this section is defined to be covariantly holomorphic
and thus satisfies:
\bea
\nabla_{\hspace{-1pt} \bar z} L^A = 0\,.
\eea
Moreover, the constrained form of the geometry implies two special properties for the two possible 
kinds of second covariant derivatives of the section. The first is that 
$\nabla_{\hspace{-1pt} \bar z} \nabla_{\hspace{-1pt} z} L^A = 
[\nabla_{\hspace{-1pt} \bar z}, \nabla_{\hspace{-1pt} z}] L^A = g_{z \bar z} L^A$.
This implies that
\bea
\nabla_{\hspace{-1pt} \bar z} \nabla_{\hspace{-1pt} z} L^A = l\, L^A \,.
\eea
The second is that $\nabla_{\hspace{-1pt}z} \nabla_{\hspace{-1pt}z} L^A = C_{z z z} g^{z \bar z} \nabla_{\hspace{-1pt}\bar z} \bar L^A$, 
where $C_{zzz}$ is covariantly holomorphic and thus satisfies $\nabla_{\hspace{-1pt}\bar z} C_{zzz} = 0$. Acting with a further derivative on this 
relation we then deduce that $\nabla_{\hspace{-1pt}\bar z} \nabla_{\hspace{-1pt}z} \nabla_{\hspace{-1pt}z} L^A = C_{z z z} g^{z \bar z} 
\nabla_{\hspace{-1pt}\bar z} \nabla_{\hspace{-1pt}\bar z}\bar L^A$. Multiplying this equation with $\nabla_{\hspace{-1pt}\bar z} \bar L^B$ 
and using the previous relation, one finally finds that
\bea
\nabla_{\hspace{-1pt}\bar z} \nabla_{\hspace{-1pt}z}{\!\!}^2 L^A \hspace{1pt} \nabla_{\hspace{-1pt}\bar z} \bar L^B 
= \nabla_{\hspace{-1pt}z}{\!\!}^2 L^B \hspace{1pt} \nabla_{\hspace{-1pt}\bar z}{\!\!}^2 \bar L^A~.
\eea

In the above parametrization in terms of the coordinates $q^\alpha = z, \bar z$, the components of the metric 
are given by:
\bea
g_{\alpha \bar \beta} = \left(\begin{matrix}
l & 0 \\
0 & l \\
\end{matrix}\right) \,.
\eea
An obvious choice for the vielbein $e_\alpha{}^\gamma$ that allows the metric to be locally trivialized as 
$g_{\alpha \bar \beta} = e_\alpha{}^\gamma \delta_{\gamma \bar \delta}\, (e^\dagger)^{\bar \delta}{}_{\bar \beta}$ is 
given by:
\bea
e_\alpha{}^\gamma = \left(\begin{matrix}
l^{1/2} & 0 \\
0 & l^{1/2} \\
\end{matrix}\right) \,.
\eea

It will be convenient for the analysis of the next sections to introduce the following dimensionless 
parameter related to the third derivatives of the section:
\bea
\beta = \frac {\nabla_{\hspace{-1pt}z}{\!\!}^3 L\, L^2}{(\nabla_{\hspace{-1pt}z} L)^3} \,.
\label{b}
\eea
It will also be useful to define the following quantity: 
\bea
\gamma = - \arg \frac {\nabla_{\hspace{-1pt} z} L}{L} \,.
\label{gamma}
\eea

\section{Models with one hyper}
\setcounter{equation}{0}

In models with just one hypermultiplet, the only possible source of potential comes 
from a gauging of the isometry with the graviphoton. This corresponds to taking 
$k_0^u = k^u$ and $P_0^x = P^x$, where $k^u$ and $P^x$ are given by eq.~(\ref{KillingPT}).
Furthermore, we can choose $L^0 = 1$. 
The resulting potential is $V = 2\,g_{uv} k^u k^v - 3\, |\vec P|^2$ and only depends on 
the three variables $q^{\hat u} = \rho, \varphi, \chi$. Its explicit form reads
\bea
V = \xi^2 \rho^{\text{--}2} \Big(f^{\text{--}1} \! - \text{\small $\frac 34$} \Big) \,.
\eea
The first derivatives $V_u \equiv \partial_u V$ are given by $V_\tau = 0$ and 
\bea
\b\b V_{\hat u} = \xi^2 \rho^{\text{--}2} \Big[\!-\! f^{\text{--}2} f_{\hat u} - 2\hspace{1pt} \rho\, \xi^{\text{--}2} V  \delta_{\hat u \rho} \Big] \,.
\eea
The second derivatives $V_{uv} \equiv \partial_u \partial_v V$ are instead given by $V_{\tau\tau} = 0$, $V_{\tau \hat v} = 0$ and
\bea
\b\b V_{\hat u \hat v} = \xi^2  \rho^{\text{--}2} \Big[\!-\! f^{\text{--}2}\! f_{\hat u \hat v}\hspace{-1pt} + 2 f^{\text{--}3}\! f_{\hat u} f_{\hat v} \hspace{-1pt} 
+ 4 \hspace{1pt} \rho^{\text{--}1}\! f^{\text{--}2} f_{(\hat u} \delta_{\hat v) \rho}
+ 6 \,\xi^{\text{--}2} V \delta_{\hat u \rho} \delta_{\hat v \rho}\Big] \,.
\eea
The gravitino mass reads:
\bea
m_{3/2}^2 = \text{\small $\frac 14$}\hspace{1pt}  \xi^2 \rho^{\text{--}2} \,.
\eea
It follows that the parameter (\ref{epsilon}) takes the value
\be
\epsilon = - 3 + 4 f^{\text{--}1} \,.
\label{epsilonh}
\ee
We see from these formulae that the strength of supersymmetry breaking is controlled by the value of $f^{\text{--}1}$. 

In order to study the possible vacua and their properties, we must first find the critical points of the scalar potential.
The value of the cosmological constant $V$ fixes the value of $f$ to be given by:
\be
f = \frac 4{3 + \epsilon} \,.
\ee
The stationarity conditions $V_u = 0$ fix instead the values of the first derivatives $f_{\hat u}$ to be
given by $f_{\hat u} = - 2\hspace{1pt} \rho f^2 \xi^{\text{--}2} V \delta_{\hat u \rho}$, and since 
$\xi^{\text{--}2} V = \frac 14 \rho^{\text{--}2} \epsilon$ this gives:
\bea
f_{\hat u} = - \frac {8\hspace{1pt} \epsilon}{(3+\epsilon)^2} \rho^{\text{--}1} \delta_{\hat u \rho} \,.
\label{stath}
\eea
Using the above relations, one can then compute the Hessian matrix $V_{uv}$ and the the physical scalar 
mass matrix $m^2_{uv} = (e^{\text{--}1})_u{}^p V_{pq} (e^{\text{--}1T})^q{}_v$ in terms of the values of 
the second derivatives $f_{\hat u \hat v}$, which are related to the parameters $\alpha_{\hat u \hat v}$ by the 
definition (\ref{a}). This is found to be given by $m^2_{\tau \tau} = 0$, $m^2_{\tau \hat v} = 0$ and
\bea
m^2_{\hat u \hat v} = \Big[\!-\! \text{\small $\frac 12$} (3 + \epsilon)^2 \alpha_{\hat u \hat v} 
- 3 \epsilon (1 - \epsilon) \delta_{\hat u \rho} \delta_{\hat v \rho} \Big]m_{3/2}^2\,.
\eea
The parameters $\alpha_{\hat u \hat v}$ can be adjusted by suitably choosing the function $f$ and therefore the 
values of its second derivatives at the point under consideration. The only constraint is that the corresponding 
function $h$ should solve the Toda equation \eqref{toda-h}. This then results in the constraint (\ref{constrainta}) 
which implies the following sum rule, after using the stationarity condition (\ref{stath}):
\bea
\delta^{\hat u \hat v} \alpha_{\hat u \hat v} = \frac {2 + 6\hspace{1pt}\epsilon^2}{(3+\epsilon)^2}\,.
\label{constrainth}
\eea

To check whether the scalar masses can all be positive, we may now compute the average of the 
three eigenvalues of the physical mass matrix:
\be
m^2 \equiv \text{\small $\frac 13$} \delta^{\hat u \hat v} m^2_{\hat u \hat v} \,.
\ee
It turns out that the value of this average mass is entirely fixed by the constraint (\ref{constrainth}) and 
is found to be:
\bea
m^2 = \Big[\!-\! \text{\small $\frac 13$} - \epsilon\Big] m^2_{3/2} \,.
\label{m2onlyhyper}
\eea
By construction $m^2$ represents an upper bound on the smallest mass eigenvalue and also a lower bound 
on the largest mass eigenvalue. 
Moreover, both of these bounds can be saturated by tuning the parameters $\alpha_{\hat u \hat v}$.
We then conclude that:
\be
\min \big\{m^2_i\big\} \le - \text{\small $\frac 13$} \,m_{3/2}^2 - V \,.
\ee
This result implies that Minkowski and de Sitter vacua are necessarily unstable. It reproduces the result derived 
in \cite{GRLS}, which looked at the average sGoldstino mass for theories with an arbitrary number of hypermultiplets.

\section{Models with one vector}
\setcounter{equation}{0}

In models with just one vector multiplet, the only possible source of potential comes 
from a Fayet-Iliopoulos term associated to constant Killing potentials for an 
arbitrary linear combination of the graviphoton and the matter vector. 
Without loss of generality, we can choose $P_0^x = 0$ and $P_1^x = \frac 12 \xi v^x$,
where $\xi$ is a constant and $v^x$ is an arbitrary unit vector. The potential is then independent of $L^0$ and 
only involves $L^1 \equiv L$. It is given by 
$V = l^{\text{--}1} |\vec P|^2 \, |\nabla_z L|^2 - 3\, |\vec P|^2 \, |L|^2$  and depends on the 
two variables $q^{\alpha} = z,\bar z$. Its explicit form reads
\bea
V \a=\a \xi^2 \Big[\,\text{\small $\frac 14$}l^{\text{--}1} |\nabla_z L|^2 - \text{\small $\frac 34$}\, |L|^2 \Big] \,.
\eea
The first derivatives $V_\alpha \equiv \nabla_{\hspace{-1pt}\alpha} V$ are found to be:
\bea
V_z \a=\a \xi^2 \Big[\,\text{\small $\frac 14$} l^{\text{--}1} \nabla_{\hspace{-1pt}z}{\!\!}^2 L\, \nabla_{\hspace{-1pt}\bar z} \bar L 
- \text{\small $\frac 12$} \nabla_{\hspace{-1pt}z} L \, \bar L \Big] \,.
\eea
The second derivatives $V_{\alpha \bar \beta} \equiv \nabla_{\hspace{-1pt}\alpha} \nabla_{\hspace{-1pt}\bar \beta} V$
are instead given by
\bea
\b\b V_{z \bar z} = \xi^2 \Big[\,\text{\small $\frac 12$} l^{\text{--}1} |\nabla_{\hspace{-1pt}z}{\!\!}^2 L|^2 
- \text{\small $\frac 12$} \big(l\, |L|^2 + |\nabla_{\hspace{-1pt}z} L|^2\big) \Big] \,,\\
\b\b V_{z z} = \xi^2 \Big[\,\text{\small $\frac 14$} l^{\text{--}1} \nabla_{\hspace{-1pt}z}{\!\!}^3 L\, \nabla_{\hspace{-1pt}\bar z} \bar L 
- \text{\small $\frac 14$} \nabla_{\hspace{-1pt} z}{\!\!}^2 L\, \bar L \Big] \,.
\eea
The gravitino mass reads:
\bea
m_{3/2}^2 = \text{\small $\frac 14$}\hspace{1pt} \xi^2 |L|^2 \,.
\eea
It follows that the parameter (\ref{epsilon}) takes the value
\be
\epsilon = - 3 + l^{\text{--}1} \bigg|\frac {\nabla_{\hspace{-1pt} z} L}{L}\bigg|^2\,.
\ee
We see from these formulae that the strength of supersymmetry breaking is controlled by the value 
of $l^{\text{--}1} |\nabla_{\hspace{-1pt} z} L/L|$.

In order to study the possible vacua and their properties, we must again first find the critical points of the scalar potential. 
The value of the cosmological constant $V$ fixes the value of 
$\nabla_{\hspace{-1pt} z} L$ to be given by
\be
\nabla_{\hspace{-1pt}z} L = \sqrt{3 + \epsilon}\,\, l^{1/2} e^{- i \gamma} L \,,
\ee
where $\gamma$ was defined in (\ref{gamma}). The stationarity conditions $V_z = 0$ fix instead the value 
of $\nabla_{\hspace{-1pt}z}{\!\!}^2 L$ to by given by
\bea
\nabla_{\hspace{-1pt}z}{\!\!}^2 L = 2\,l\,e^{\text{--}2 i \gamma} L \,.
\eea
Using the above relations, one can then compute the Hessian matrix $V_{\alpha \bar \beta}$ and 
the physical scalar mass matrix $m^2_{\alpha \bar \beta} = (e^{\text{--}1})_\alpha{}^\epsilon 
V_{\epsilon \bar \delta} (e^{\text{--}1\dagger})^{\bar \delta}{}_{\bar \beta}$ in terms of the value of 
$\nabla_{\hspace{-1pt}z}{\!\!}^3 L$, which is related to the parameter $\beta$ by the definition (\ref{b}). 
This gives:
\bea
\b\b m^2_{z \bar z} = \Big[\!-\! 2\,\epsilon\Big] m^2_{3/2} \,, \\
\b\b m^2_{z z} = \Big[(3 + \epsilon)^2 \beta -2\Big] m^2_{3/2} e^{\text{--}2 i \gamma} \,.
\eea
The parameter $\beta$ can be adjusted to any desired value by changing the function $l$ and therefore
the value of its second derivative at the point under consideration, without any constraint.

To check whether the scalar masses can all be positive, we can now compute the average of the 
two eigenvalues of the physical mass matrix:
\be
m^2 \equiv \text{\small $\frac 12$} \delta^{\alpha \bar \beta} m^2_{\alpha \bar \beta} \,.
\ee
We see that this is completely fixed, independently of the parameter $\beta$, and reads
\bea
m^2 = \Big[\!-\! 2\,\epsilon \Big] m^2_{3/2} \,.
\label{m2onlyvec}
\eea
By construction $m^2$ represents an upper bound on the smallest mass eigenvalue and also a lower bound 
on the largest mass eigenvalue. 
Moreover, both of these bounds can be saturated by tuning the parameter $\beta$.
From this we deduce that
\be
\min \big\{m^2_i\big\} \le -2\, V \,.
\ee
This result implies that de Sitter vacua are necessarily unstable. It reproduces the result 
derived in \cite{MANY}, which looked at the average sGoldstino mass for theories 
involving an arbitrary number of Abelian vector multiplets.

\section{Models with one hyper and one vector}
\setcounter{equation}{0}

In models with one hypermultiplet and one vector multiplet, the source of potential 
comes from a gauging of the hypermultiplet isometry with a linear combination of the graviphoton
and the matter vector. Without loss of generality, we can choose $k_0^u = 0$, $k_1^u = k^u$ and 
$P_0^x = 0$, $P_1^x = P^x$, where $k^u$ and $P^x$ are given by eq.~(\ref{KillingPT}). Once again,
the potential involves $L^1 \equiv L$ but not $L^0$. It is given by
$V = 2\,g_{uv} k^u k^v |L|^2 + l^{\text{--}1} |\vec P|^2 \, |\nabla_z L|^2 - 3\, |\vec P|^2 \, |L|^2$
and depends only on the five variables $q{}^{\hat I} = \rho, \varphi, \chi,z,\bar z$. Its explicit expression 
is
\bea
V = \xi^2 \rho^{\text{--}2} \Big[\Big(f^{\text{--}1}\! - \text{\small $\frac 34$} \Big) |L|^2 + \text{\small $\frac14$}\, l^{\text{--}1}\, |\nabla_z L|^2 \Big] \,.
\eea
The first derivatives $V_I$ are given by $V_\tau = 0$ and 
\bea
\b\b V_{\hat u} = \xi^2 \rho^{\text{--}2} \Big[\!-\! f^{\text{--}2} f_{\hat u} |L|^2 - 2 \rho\, \xi^{\text{--}2} V \delta_{\hat u \rho} \Big] \,, \\
\b\b V_z = \xi^2  \rho^{\text{--}2} \Big[\text{\small $\frac 14$} l^{\text{--}1} \nabla_{\hspace{-1pt}z}{\!\!}^2 L\, \nabla_{\hspace{-1pt} \bar z} \bar L
- \text{\small $\frac 12$} \big(1- 2 f^{\text{--}1} \big) \nabla_z L \, \bar L \Big] \,.
\eea
The second derivatives $V_{I\bar J}$, defined to be ordinary and covariant in the hyper and vector sectors, respectively, 
are instead given by $V_{\tau\tau} = 0$, $V_{\tau \hat J} = 0$ and
\bea
\b\b V_{\hat u \hat v} = \xi^2 \rho^{\text{--}2} \Big[\big(\!-\! f^{\text{--}2} f_{\hat u \hat v} + 2 f^{\text{--}3}\! f_{\hat u} f_{\hat v} 
+ 4\hspace{1pt} \rho^{\text{--}1} f^{\text{--}2} f_{(\hat u} \delta_{\hat v) \rho} \big) |L|^2 
+ 6 \,\xi^{\text{--}2} V \delta_{\hat u \rho} \delta_{\hat v \rho}\Big] \,, \\
\b\b V_{z \bar z} = \xi^2 \rho^{\text{--}2} \Big[\text{\small $\frac 12$} l^{\text{--}1} |\nabla_{\hspace{-1pt}z}{\!\!}^2 L|^2 
- \text{\small $\frac 12$} \big(1- 2 f^{\text{--}1} \big) \big(l\, |L|^2+ |\nabla_{\hspace{-1pt}z} L|^2\big) \Big] \,,\\
\b\b V_{z z} = \xi^2 \rho^{\text{--}2} \Big[\text{\small $\frac 14$} l^{\text{--}1} \nabla_{\hspace{-1pt}z}{\!\!}^3 L\, \nabla_{\hspace{-1pt}\bar z} \bar L 
- \text{\small $\frac 14$} \big(1- 4 f^{\text{--}1} \big) \nabla_{\hspace{-1pt} z}{\!\!}^2 L\, \bar L \Big] \,, \\
\b\b V_{\hat u z} = \xi^2 \rho^{\text{--}2} \Big[\!-\! f^{\text{--}2} f_{\hat u} \nabla_z L\, \bar L + \rho^{\text{--}1}
\big(\big(1- 2 f^{\text{--}1} \big) \nabla_{\hspace{-1pt}z} L \, \bar L 
- \text{\small $\frac 12$} l^{\text{--}1} \nabla_{\hspace{-1pt}z}{\!\!}^2 L \, \nabla_{\hspace{-1pt}\bar z} \bar L \big) 
\delta_{\hat u \rho} \Big] \,.
\eea
The gravitino mass reads:
\bea
m_{3/2}^2 = \text{\small $\frac 14$}\hspace{1pt} \xi^2 \rho^{\text{--}2} |L|^2 \,.
\eea
It follows that the parameter (\ref{epsilon}) is given by
\be
\epsilon = - 3 + 4 f^{\text{--}1} + l^{\text{--}1} \bigg|\frac {\nabla_{\hspace{-1pt} z} L}{L} \bigg|\,.
\ee
We see from these formulae that the strength of supersymmetry breaking originating from the hyper and vector sectors 
is controlled by the expectation values of $f^{\text{--}1}$ and $l^{\text{--}1} |\nabla_{\hspace{-1pt} z} L/L|$, respectively. 
The relative importance of these two contributions is most conveniently parametrized by an angle $\theta$, defined as:
\bea
\tan \theta = \text{\small $\frac 12$} \sqrt{\frac fl}\, \bigg|\frac {\nabla_z L}{L}\bigg| \,.
\eea

In order to study the possible vacua and their properties, we must again first find the critical points of the scalar potential.
The values of the cosmological constant $V$ and the angle 
$\theta$ fix the values of $f$ and $\nabla_{\hspace{-1pt} z} L$ to:
\bea
\b\b f = \frac 4{3 +\epsilon} \cos^{\text{--}2}\! \theta 
\label{cos} \,, \\
\b\b \nabla_{\hspace{-1pt}z} L = \sqrt{3 + \epsilon}\,\, l^{1/2} \sin \theta\, e^{- i \gamma} L 
\label{sin} \,,
\eea
where we recall again that $\gamma$ was defined in (\ref{gamma}). 
The stationarity conditions $V_I = 0$ fix instead the values of $f_{\hat u}$ and 
$\nabla_{\hspace{-1pt}z}{\!\!}^2 L$ to by given by $f_{\hat u} = - 2 \rho f^2 \xi^{\text{--}2}|L|^{\text{--}2} V \delta_{\hat u \rho}$
and $\nabla_{\hspace{-1pt}z}{\!\!}^2 L = 2\,l\, \big(1- 2 f^{\text{--}1} \big)  e^{\text{--} 2 i \gamma} L$, 
and since $\xi^{\text{--}2} |L|^{\text{--}2}V = \frac 14 \rho^{\text{--}2} \epsilon$ this gives:
\bea
\b\b f_{\hat u} = - \frac {8\hspace{1pt} \epsilon}{(3+\epsilon)^2} \cos^{\text{--}4}\! \theta \, \rho^{\text{--}1} \delta_{\hat u \rho} 
\label{stathv1}\,, \\
\b\b \nabla_{\hspace{-1pt}z}{\!\!}^2 L = -\hspace{1pt} l \hspace{1pt} \big((3 + \epsilon) \cos^2\! \theta - 2\big)e^{\text{--}2 i \gamma} L 
\label{stathv2} \,.
\eea
Using the above relations, one can then compute the Hessian matrix $V_{I \bar J}$ and 
the physical scalar mass matrix $m^2_{I \bar J} = (e^{\text{--}1})_I{}^P V_{P \bar Q} (e^{\text{--}1\dagger})^{\bar Q}{}_{\bar J}$
in terms of the values of $f_{\hat u \hat v}$ and $\nabla_{\hspace{-1pt}z}{\!\!}^3 L$, which are related to the parameters 
$\alpha_{\hat u \hat v}$ and $\beta$ by the definitions (\ref{a}) and (\ref{b}). This is found to be given by
$m^2_{\tau \tau} = 0$, $m^2_{\tau \hat v} = 0$, $m^2_{\tau \alpha} = 0$ and
\bea
\b\b m^2_{\hat u \hat v} = \Big[\!-\! \text{\small $\frac 12$}\big(3 + \epsilon\big)^2 \cos^4\! \theta \,\alpha_{\hat u \hat v}
- \epsilon \big((3 + \epsilon) \cos^2\! \theta - 4\,\epsilon\big) \delta_{\hat u \rho} \delta_{\hat v \rho} \Big] m^2_{3/2}\,, \\
\b\b m^2_{z \bar z} = \Big[\big((3 + \epsilon) \cos^2\! \theta - 2\big) \big((3 + \epsilon) \cos^2\! \theta + \epsilon\big) \Big] m^2_{3/2} \,, \\
\b\b m^2_{z z} = \Big[(3 + \epsilon)^2 \! \sin^4\! \theta \, \beta - \big((3 + \epsilon) \cos^2\! \theta - 1\big) \big((3 + \epsilon) \cos^2\! \theta - 2\big)\Big] 
m^2_{3/2} e^{\text{--}2 i \gamma}\,, \\
\b\b m^2_{\hat u z} = \Big[\sqrt{2}\,\epsilon\, (3 + \epsilon) \cos \theta \sin \theta\, \delta_{\hat u \rho} \Big] m^2_{3/2} e^{\text{--}i \gamma} \,.
\eea
The parameters $\alpha_{\hat u \hat v}$ can be adjusted by suitably choosing the function $f$ and therefore the 
values of its second derivatives at the point under consideration. The only constraint comes from the Toda equation 
for $h$ \eqref{toda-h}, which can be rewritten as in (\ref{constrainta}). After using the stationarity condition (\ref{stathv1}) 
this in turn implies the following sum rule:
\bea
\delta^{\hat u \hat v} \alpha_{\hat u \hat v} = \frac {- 2 (3+\epsilon)^2 + 4 (3 + \epsilon) (3 + 2 \epsilon) \cos^{\text{--}2}\!\theta
- 8 (2 + 3 \epsilon) \cos^{\text{--}4}\! \theta} {(3+\epsilon)^2}\,.
\label{constrainthv}
\eea
The parameter $\beta$ can instead be adjusted by suitably choosing the function $l$ and the associated section, 
and is completely arbitrary. 
Finally, the angle $\theta$ can be adjusted by suitably choosing the relative overall sizes of $f$ and $l$.

To check whether the scalar masses can all be positive, we can now study the two-dimensional matrix obtained 
by averaging over the three non-trivial directions in the hyper sector and the two directions in the vector sector.
This real symmetric matrix takes the form
\bea
m^2 \equiv \left(\,\begin{matrix}
m^2_{\rm hh} \!&\! m^2_{\rm hv} \smallskip\ \\
m^2_{\rm hv}  \!&\! m^2_{\rm vv} \\
\end{matrix} \! \right) \,,
\label{m2avr}
\eea
and its three independent elements are defined by
\be
m^2_{\rm hh} \equiv \text{\small $\frac 13$} \delta^{\hat u \hat v} m^2_{\hat u \hat v} \,,\;\;
m^2_{\rm vv} \equiv \text{\small $\frac 12$} \delta^{\alpha \bar \beta} m^2_{\alpha \bar \beta} \,,\;\;
m^2_{\rm hv} \equiv  \sqrt{\text{\small $\frac 16$} \delta^{\hat u \hat v}  \delta^{\alpha \bar \beta} m^2_{\hat u \alpha} m^2_{\hat v \bar \beta}} \,.
\ee
It turns out that the values of these average entries of the mass matrix are 
entirely fixed, as in the previous cases, and are found to be 
\bea
\b\b m^2_{\rm hh} = \Big[\text{\small $\frac 13$} \big((3 + \epsilon) \cos^2\! \theta - (2 + \epsilon) \big) 
\big((3 + \epsilon) \cos^2\! \theta - 4 (1 + \epsilon) \big) \Big]m^2_{3/2}
\label{m2hh} \,, \\
\b\b m^2_{\rm vv} = \Big[\big((3 + \epsilon) \cos^2\! \theta - 2\big) \big((3 + \epsilon) \cos^2\! \theta + \epsilon\big) \Big] m^2_{3/2} 
\label{m2vv}\,, \\[-1mm]
\b\b m^2_{\rm hv} = \Big[\text{\small $\sqrt{\frac 23}$}\hspace{1pt} \epsilon \hspace{1pt} (3 + \epsilon) \cos \theta \sin \theta \Big] m^2_{3/2} 
\label{m2hv}\,.
\eea
Note that in the hyper sector one correctly recovers $m^2_{\rm hh} \to - \frac 13 (1 + 3\epsilon) \, m_{3/2}^2$ 
when $\theta \to 0$, matching the case with only one hypermultiplet (\ref{m2onlyhyper}), and one finds instead 
$m^2_{\rm hh} \to \frac 43 (1 + \epsilon) (2 + \epsilon) m_{3/2}^2$ when $\theta \to \frac {\pi}2$.
Similarly, in the vector sector one correctly recovers $m^2_{\rm vv} \to - 2 \epsilon \, m_{3/2}^2$ when 
$\theta \to \frac {\pi}2$, matching the case with only one vector multiplet (\ref{m2onlyvec}), and one finds 
instead $m^2_{\rm vv} \to (1 + \epsilon) (3 + 2 \epsilon)\, m_{3/2}^2$ when $\theta \to 0$. Finally, for the mixing 
between the two sectors, one finds $m^2_{\rm hv} \to 0$ when either $\theta \to 0$ or $\theta \to \frac {\pi}2$.

To get the sharpest possible bounds on the mass eigenvalues for the case we consider here, it is natural to consider the two eigenvalues 
of the two-dimensional averaged mass matrix (\ref{m2avr}). These are given by
\be
m^2_{\rm \pm} = \text{\small $\frac 12$} \big(m^2_{\rm hh} + m^2_{\rm vv}\big)
\pm \sqrt{\text{\small $\frac 14$} \big(m^2_{\rm hh} - m^2_{\rm vv} \big)^2 + m^4_{\rm hv}} \,.
\ee
Using the results (\ref{m2hh}), (\ref{m2vv}) and (\ref{m2hv}), these two eigenvalues are found to be of the form
\bea
m^2_{\pm} = \Big[X \pm \sqrt{Y} \Big] m^2_{3/2}\,,
\label{m2+-}
\eea
where:
\bea
\b\b X = \text{\small $\frac 23$} (3 + \epsilon)^2 \cos^4\! \theta - \text{\small $\frac 13$} (3 + \epsilon) (6 + \epsilon) \cos^2\! \theta
+ \text{\small $\frac 13$} (4 + 3 \epsilon + 2 \epsilon^2) \,, \label{X} \\
\b\b Y = \text{\small $\frac 19$} (3 + \epsilon)^4 \cos^8\! \theta +  \text{\small $\frac 89$} (3 + \epsilon)^3 \epsilon \cos^6\! \theta
-  \text{\small $\frac 29$}(3 + \epsilon)^2 (4 + 9 \epsilon - 3\epsilon^2) \cos^4\! \theta \nn \\ 
\b\b \hspace{23pt} -\, \text{\small $\frac 29$} (3 + \epsilon) \epsilon (16 + 27 \epsilon + 5 \epsilon^2) \cos^2\! \theta
+  \text{\small $\frac 19$} (4 + 9 \epsilon + 2 \epsilon^2)^2 \,. \label{Y}
\eea
We can now study the behavior of $m^2_\pm$ as functions of $\theta$ for given value of $\epsilon$. 
For $m^2_-$, one finds a local maximum for some finite value of $\theta$, with a magnitude that is negative 
or positive depending on whether $\epsilon$ is smaller or larger than a certain critical value, as illustrated 
in figure 1. More precisely, one finds the following results. For $\epsilon=0$, $m^2_-$ is everywhere negative, 
except at $\theta = \arccos \raisebox{-3pt}{$\sqrt{\;\; \raisebox{13pt}{$$}}$}\hspace{-7pt} \frac 23 \simeq 0.62$
where it vanishes. This implies that Minkowski vacua ($V=0$) can be at most loosely metastable. For $\epsilon \in [0, 2.17[\,$, 
$m^2_-$ is everywhere negative, and de Sitter vacua with $V \lsim 2.17\, m_{3/2}^2$ are thus necessarily unstable. 
For $\epsilon \simeq 2.17$, $m^2_-$ is everywhere negative except at $\theta \simeq 0.65$ where it vanishes,
and de Sitter vacua with $V \simeq 2.17\, m_{3/2}^2$ can thus be loosely metastable. Finally for $\epsilon \gsim 2.17$, 
$m^2_-$ becomes positive for a finite range of values of $\theta$, centered around a value between $0.65$ and 
$\frac {\pi}4 \simeq 0.79$ where the maximum occurs. Therefore de Sitter vacua with $V > 2.17\, m_{3/2}^2$ can be 
genuinely metastable. For $m^2_+$, one finds a local minimum for some finite value of $\theta$, with a magnitude that 
is always positive.

\begin{figure}[h]
\vskip 20pt
\begin{center}
\includegraphics[width=0.6\textwidth]{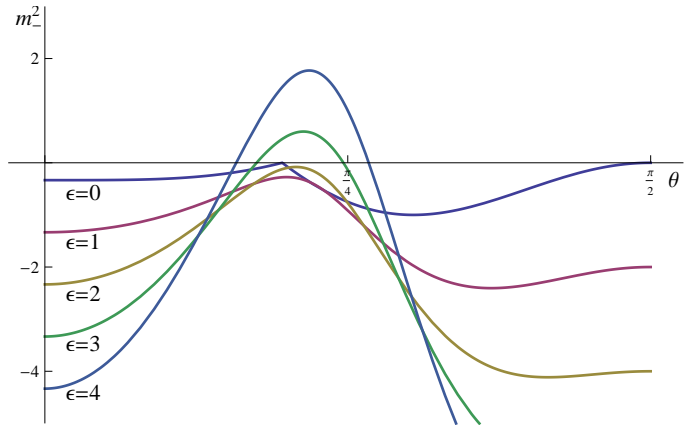}
\caption{Plot of $m^2_-$ as a function of $\theta$ for various values of $\epsilon$, in units of $m^2_{3/2}$.}
\end{center}
\end{figure}

By construction $m^2_-$ represents an upper bound on the smallest mass eigenvalue and $m^2_+$ a lower bound on the largest
mass eigenvalue, for given values of $\theta$ and $\epsilon$. From these quantities one can then derive two other bounds 
that only depend on $\epsilon$ by suitably extremizing $m^2_-$ and $m^2_+$ over $\theta$, for fixed $m^2_{3/2}$ and $V$. More 
precisely, we can compute:
\bea
\b\b m^2_{\rm up} \equiv \max_\theta \big\{m^2_-\big\} \,, \\
\b\b m^2_{\rm low} \equiv \min_\theta \big\{m^2_+\big\} \,.
\eea
One can then see that $\min\{m^2_i\} \le m^2_- \le m^2_{\rm up}$ and $\max\{m^2_i\} \ge m^2_+ \ge m^2_{\rm low}$.
Moreover, any of these bounds can be saturated by tuning the parameters $\alpha_{\hat u \hat v}$, $\beta$ and $\theta$.
We can now study the behavior of $m^2_{\rm up,low}$ as functions of $\epsilon$. For $m^2_{\rm up}$, one 
finds a non-monotonic function that first decreases and is negative and then increases and becomes positive, 
as shown in figure 2. For $m^2_{\rm low}$, one finds instead a monotonically increasing function that is always positive. 

\begin{figure}[h]
\vskip 20pt
\begin{center}
\includegraphics[width=0.6\textwidth]{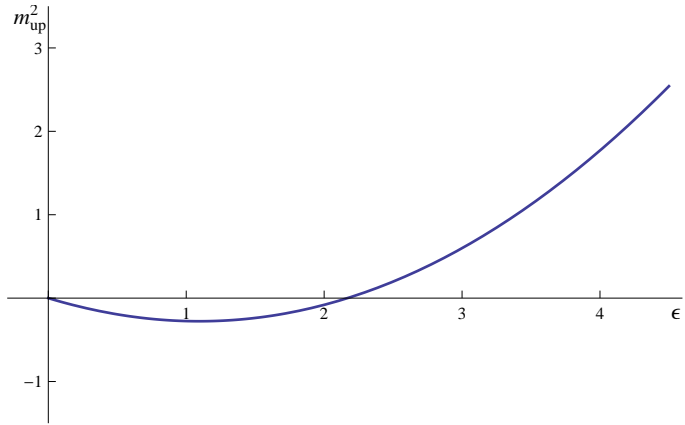}
\caption{Plot of $m^2_{\rm up}$ as a function of $\epsilon$, in units of $m^2_{3/2}$.}
\end{center}
\end{figure}

To summarize, we see that when $V$ is positive and small compared to $m^2_{3/2}$, as required for particle phenomenology, 
the vacuum is necessarily unstable, but when $V$ is positive and large compared to $m^2_{3/2}$, as could be desirable 
for inflation, the vacuum can be metastable. In these two limits of small and large cosmological constant, one can actually 
derive some simpler result for the bound on the smallest mass eigenvalue. This will also allow us to describe the general 
case of intermediate cosmological constant in a simpler, qualitatively way. 

In the limit where $\epsilon$ is small, which corresponds to a situation where the cosmological constant is 
made small through a tuning of parameters, the entries of the two-dimensional averaged mass matrix are 
given by:
\bea
\b\b m^2_{\rm hh} \simeq \Big[\Big(\text{\small $\frac 83$} - 6 \cos^2\! \theta + 3 \cos^4\! \theta \Big) 
+ \Big(4 - 7 \cos^2\! \theta + 2 \cos^4\! \theta \Big) \epsilon \Big]m^2_{3/2} \,, \label{mhhsmall} \\
\b\b m^2_{\rm vv} \simeq \Big[\Big(\!- 6 \cos^2\! \theta + 9 \cos^4\! \theta\Big) 
+ \Big(\!-2 + \cos^2\! \theta + 6 \cos^4\! \theta\Big) \epsilon\Big] m^2_{3/2} \,, \label{mvvsmall} \\
\b\b m^2_{\rm hv} \simeq \Big[\Big(0\Big) + \Big(\sqrt{6} \cos \theta \sin \theta\Big) \epsilon \Big] m^2_{3/2}\,. \label{mhvsmall}
\eea
When $\epsilon = 0$, the two eigenvalues $m^2_\pm$ of (\ref{m2avr}) take degenerate and vanishing 
extremal values $m^2_{\rm up,low}$ for the same angle $\theta = \theta_0$ given by 
$\theta_0 = \arccos \raisebox{-3pt}{$\sqrt{\;\; \raisebox{13pt}{$$}}$}\hspace{-7pt} \frac 23 \simeq 0.62$.
When $\epsilon \neq 0$, the two eigenvalues $m^2_\pm$ of (\ref{m2avr}) instead take non-degenerate and 
non-vanishing extremal values $m^2_{\rm up,low}$ for two slightly different angles 
$\theta = \theta_0 + \Delta \theta_\pm$. To compute the approximate values of $m^2_{\rm up,low}$ 
for small $\epsilon$, one can then further expand the entries (\ref{mhhsmall}), (\ref{mvvsmall}) and (\ref{mhvsmall}) 
at $\theta = \theta_0 + \Delta \theta_\pm$ up to first order in $\Delta \theta_\pm = \kappa_\pm\, \epsilon$, then 
compute the eigenvalues and finally extremize them with respect to the parameter $\kappa_\pm$. Proceeding 
in this way one finds:
\bea
\b\b m^2_{\rm up} \simeq - \text{\small $\frac 12$} \epsilon\, m^2_{3/2} \,, \\
\b\b m^2_{\rm low} \simeq \text{\small $\frac 32$}  \epsilon\, m^2_{3/2} \,.
\eea
It follows that in this regime:
\bea
\min \big\{m^2_i \big\} \lsim - \text{\small $\frac 12$}\, V \,.
\eea
This quantifies the extent to which de Sitter vacua with a small cosmological constant are unstable 
in this class of models.

In the limit where $\epsilon$ is large, which includes the situation where the main contribution to the 
vacuum energy is non-gravitational, the entries of the two-dimensional averaged mass matrix are 
instead given by:
\bea
\b\b m^2_{\rm hh} \simeq \Big[\text{\small $\frac 13$} \big(\cos^2\! \theta - 1 \big) \big(\cos^2\! \theta - 4 \big) \Big] \epsilon^2 m^2_{3/2} \,, \\
\b\b m^2_{\rm vv} \simeq \Big[\cos^2\! \theta \big(\cos^2\! \theta + 1\big) \Big] \epsilon^2 m^2_{3/2} \,, \\[-1mm]
\b\b m^2_{\rm hv} \simeq \Big[\text{\small $\sqrt{\frac 23}$}\hspace{1pt} \cos \theta \sin \theta \Big] \epsilon^2 m^2_{3/2} \,.
\eea
After computing $m^2_{\pm}$, one can check that $m^2_-$ develops a maximum for $\theta \simeq \frac {\pi}4 \simeq 0.79$ 
and $m^2_+$ develops a minimum for $\theta \simeq 0.89$, with the values
\bea
\b\b m^2_{\rm up} \simeq \text{\small $\frac 14$} \epsilon^2 m_{3/2}^2 \,, \\
\b\b m^2_{\rm low} \simeq 1.05\, \epsilon^2 m_{3/2}^2 \,.
\eea
It follows that in this regime:
\bea
\min \big\{m^2_i \big\} \lsim \text{\small $\frac 14$}\,\frac {V^2}{m^2_{3/2}} \,.
\eea
This quantifies the extent to which de Sitter vacua with a large cosmological constant are stable 
in this class of models.

In more general situations where $\epsilon$ is neither very large nor very small, one has in principle to use the 
exact entries of the two-dimensional averaged mass matrix. The extrema $m^2_{\rm up}$ and $m^2_{\rm low}$
must then be computed by extremizing the corresponding complicated expressions (\ref{m2+-}) for $m^2_-$ and $m^2_+$.
However, it turns out that the exact results for these bounds are reasonably well approximated for any value of 
$\epsilon$ by just adding up the behaviors that we derived for small and large values of $\epsilon$. One can 
in fact write:
\bea
\b\b m^2_{\rm up} < \Big(\!-\! \text{\small $\frac 12$} \epsilon + \text{\small $\frac 14$} \epsilon^2 \Big) m^2_{3/2} \,, \\
\b\b m^2_{\rm low} > \Big(\text{\small $\frac 32$} \epsilon + 1.05\, \epsilon^2\Big)m^2_{3/2} \,.
\eea
It follows in particular that:
\bea
\min \big\{m^2_i \big\} < -  \text{\small $\frac 12$} \, V +  \text{\small $\frac 14$} \, \frac {V^2}{m^2_{3/2}} \,.
\eea
This quantifies in a simpler but weaker way the situation for de Sitter vacua with an 
intermediate cosmological constant in this class of models. Note that this alternative 
bound can no longer be strictly saturated, and in relation to this the transition between positive 
and negative values on the right-hand side occurs at $\epsilon=2$ rather than $\epsilon \simeq 2.17$.

\section{Rigid limit}
\setcounter{equation}{0}

Let us briefly study what happens in the rigid limit, when gravitational effects are negligible, in order to compare our 
results with those of \cite{LSS}. In this decoupling limit $m^2_{3/2} \to 0$ while $V$ stays finite but is dominated 
by non-gravitational effects. One is then in the situation where $\epsilon \to +\infty$, and the entries of the mass matrix 
simplify. Moreover, all these entries take the form of a finite coefficient depending on the angle $\theta$ times 
the squared energy scale $\epsilon^2 m^2_{3/2}$, which stays finite in the limit. More precisely, one finds
\bea
\b\b m^2_{\hat u \hat v} = \Big[\!-\! \text{\small $\frac 12$} \cot^2\! \theta \,\alpha_{\hat u \hat v}
+ \big(7 + 3 \cot^2\! \theta + 4 \tan^2\! \theta\big) \delta_{\hat u \rho} \delta_{\hat v \rho} \Big] m^2_A\,,  \label{muvrigid} \\
\b\b m^2_{z \bar z} = \Big[1 + 2 \cot^2\! \theta \Big] m^2_A \,, \label{mzzbrigid} \\
\b\b m^2_{z z} = \Big[\tan^2\! \theta\, \beta - \cot^2\! \theta \Big] m^2_A e^{\text{--}2 i \gamma} \,, \label{mzzrigid}\\
\b\b m^2_{\hat u z} = \Big[\sqrt{2}\,\big(\cot \theta + \tan \theta\big)\, \delta_{\hat u z} \Big] m^2_A e^{\text{--} i \gamma} \label{muzrigid}\,,
\eea
where $m^2_A = \cos^2\! \theta \sin^2\! \theta \, \epsilon^2 m^2_{3/2}$, which using the relations 
(\ref{cos}) and (\ref{sin}) is recognized to correspond to the vector mass in the rigid limit, namely
\bea
m^2_A = \xi^2 \rho^{\text{--}2} \! f^{\text{--}1} l^{\text{--}1} |\nabla_{\hspace{-1pt}z} L|^2\,.
\eea
Notice also that in the rigid limit the constraint that the parameters $\alpha_{\hat u \hat v}$ must satisfy 
simplifies to 
\bea
\delta^{\hat u \hat v} \alpha_{\hat u \hat v} = 6 + 8 \tan^2\! \theta \,.
\eea
The averages of the above blocks are then found to be:
\bea
\b\b m^2_{\rm hh} = \Big(1 + \text{\small $\frac 43$} \tan^2\! \theta \Big) m^2_A \,,  \label{m2hhrigidlimit} \\
\b\b m^2_{\rm vv} = \Big(1 + 2 \cot^2\! \theta \Big) m^2_A \,, \label{m2vvrigidlimit} \\[-1.2mm]
\b\b m^2_{\rm hv} = \text{\small $\sqrt{\frac 23}$} \Big(\tan \theta + \cot \theta \Big) m^2_A \,. \label{m2hvrigidlimit} 
\eea
Notice finally that in the rigid limit it is more meaningful to study the extrema of $m^2_\pm$ with respect to $\theta$ at 
fixed $m^2_A$. By doing so, one finds that the extrema of $m^2_\pm$ both occur at $\theta = \frac {\pi}4$, with 
values $m^2_A$ and $\frac {13}3 m^2_A$. This implies in particular that
\bea
\min \big\{m^2_i \big\} \le m^2_A \,.
\eea

At this point, we can compare the above formulae with the results derived in \cite{LSS} in rigid supersymmetry,
where the Przanowski-Tod quaternionic-K\"ahler manifold reduces to a Gibbons-Hawking hyper-K\"ahler manifold \cite{GH1,GH2,HiKLR}
and the local special-K\"ahler manifold reduces to a global special-K\"ahler manifold. We see that 
the averaged masses (\ref{m2hhrigidlimit}), (\ref{m2vvrigidlimit}) and (\ref{m2hvrigidlimit}) exactly match the corresponding 
results in \cite{LSS}. This is expected, since these quantities are, by definition, independent of any parameter and 
any coordinate choice. On the other hand, the non-averaged masses (\ref{muvrigid}), (\ref{mzzbrigid}), (\ref{mzzrigid}) 
and (\ref{muzrigid}) can be compared with those in \cite{LSS} only after taking into account the differences in the 
coordinates used here and in \cite{LSS} to describe the scalar manifold and the associated parameters. 

In the hypermultiplet sector, the parameters $\alpha_{\hat u \hat v}$ defined here must map to the parameters 
$a_{ij}$ defined in \cite{LSS}. However, the coordinates $\rho, \varphi, \chi, \tau$ used here are related in a non-trivial 
way to the coordinates $x_i,t$ used in \cite{LSS}. By comparing the form of the metric, the Killing vector and the 
Killing potentials for the two spaces in the Ricci-flat limit where they should coincide, one can see that $\rho$ 
is related to $|\vec x|^{\text{--}1/2}$, while $\varphi$ and $\chi$ are related to the two angular variables $u,v$ 
describing the orientation of the vector $\vec x/|\vec x|$, and finally $\tau$ is related to $t$. 
Moreover, the function $f$ in the Przanowski-Tod metric is directly related to the function $f$ in the Gibbons-Hawking 
metric by a simple rescaling involving two powers of the radial coordinate. The precise relation between 
the complete $\alpha_{\hat u \hat v}$ and the complete $a_{ij}$ is then not totally straightforward to determine,
due to this non-trivial change of coordinates involving the Planck scale (see \cite{AADT} for a related discussion
in a specific example). For this reason, we will not attempt to compare more 
explicitly the non-averaged mass matrix in this sector with the results of \cite{LSS}.

In the vector multiplet sector, the parameter $\beta$ defined here must map to the parameter $b$ defined in \cite{LSS}. 
Moreover, the coordinate $z$ used here maps to the coordinate $z$ used in \cite{LSS}. Similarly, the function 
$l$ used here maps to the function $l$ in \cite{LSS}. The precise relation between $\beta$ and $b$ is thus 
straightforward to determine. Using special coordinates and taking 
the rigid limit (in which the K\"ahler connection drops out from the covariant derivatives) the symplectic 
section simply reads $L = z$, and one finds that $\beta = - b + 3 \cot^4 \theta$. Taking into account this relation, 
the structure of the non-averaged mass 
matrix in this sector then precisely matches the results of \cite{LSS} in the rigid limit, as it should.

\section{Examples}
\setcounter{equation}{0}

In order to illustrate the general statements of the previous sections, let us now study a class of more explicit 
examples and compute the full spectrum of their mass eigenvalues. For concreteness and simplicity, we shall focus on
a family of solutions where the eigenvalues of the mass matrix can be computed analytically. More precisely, 
let us keep $\theta$ and $\epsilon$ arbitrary, but take all the real parameters $\alpha_{\hat u \hat v}$ to be 
correlated and controlled by a single real parameter $x$ and similarly choose the complex parameter $\beta$ 
to be controlled by a single real parameter $y$:
\bea
\b\b \alpha_{\rho \rho} = \frac {- 2 (1 + x)(3+\epsilon)^2 + 4 (3 + \epsilon) (3 + 2 \epsilon) \cos^{\text{--}2}\!\theta
- 8 (2 + 3 \epsilon) \cos^{\text{--}4}\! \theta} {(3+\epsilon)^2} \,, \\[-1mm]
\b\b \alpha_{\varphi\varphi} = x \,,\;\; \alpha_{\chi\chi} = x \,,\;\; \alpha_{\rho_\varphi} = 0 \,,\;\; \alpha_{\rho \chi} = 0 \,,\;\; \alpha_{\varphi \chi} = 0\,, \\[1mm]
\b\b \beta = y \,.
\eea
For simplicity we shall also set $\gamma = 0$, as this does not affect the eigenvalues. With this choice of parameters the 
mass matrix takes the form
\bea
m^2_{I \bar J} = \lambda_{I \bar J}\, m^2_{3/2} \,,
\eea
where the non-trivial entries of the matrix $\lambda_{I \bar J}$ are given by
\bea
\b\b \lambda_{\rho\rho} = (3 +  \epsilon)^2 (1 + x) \cos^4\! \theta - (3 + \epsilon) (6 + 5 \epsilon) \cos^2\! \theta + 4 (1 + \epsilon) (2 + \epsilon) \,, \\
\b\b \lambda_{\varphi\varphi} = - \text{\small $\frac 12$} (3 + \epsilon)^2 x \cos^4\! \theta \,,\;\;
\lambda_{\chi\chi} = - \text{\small $\frac 12$} (3 + \epsilon)^2 x \cos^4\! \theta \,, \\[0mm]
\b\b \lambda_{\rho \varphi} = 0 \,,\;\; \lambda_{\rho \chi} = 0 \,,\;\; \lambda_{\varphi \chi} = 0 \,, \\[1mm]
\b\b \lambda_{z \bar z} = \big((3 + \epsilon) \cos^2\! \theta - 2\big) \big((3 + \epsilon) \cos^2\! \theta + \epsilon\big) \,, \\[1mm]
\b\b \lambda_{z z} = (3 + \epsilon)^2 \! \sin^4\! \theta \,y - \big((3 + \epsilon) \cos^2\! \theta - 1\big) \big((3 + \epsilon) \cos^2\! \theta - 2\big) \,, \\[1mm]
\b\b \lambda_{\rho z} = \sqrt{2}\,\epsilon\, (3 + \epsilon) \cos \theta \sin \theta \,,\;\; \lambda_{\varphi z} = 0 \,,\;\; \lambda_{\chi z} = 0 \,.
\eea
The five non-trivial masses are then given in terms of  the eigenvalues $\lambda_i$ of 
the non-trivial block $\lambda_{\hat I \hat{\bar J}}$ of this matrix:
\bea
m^2_i = \lambda_i\, m^2_{3/2} \,.
\eea
In this simple four-parameter special family of vacua, the eigenvalues $\lambda_i$ can be computed analytically 
as functions of the parameters $\theta$, $\epsilon$, $x$ and $y$ characterizing the scalar geometry in the vicinity
of the vacuum point. They are found to be
\bea
\b\b \lambda_{1,2} = - \text{\small $\frac 12$} (3 + \epsilon)^2 x \cos^4\! \theta \,, \\
\b\b \lambda_3 = - (3 + \epsilon)^2 y \sin^4\! \theta 
+ \big((3 + \epsilon) \cos^2\! \theta -2 \big) \big(2 (3 + \epsilon) \cos^2\! \theta -1 + \epsilon \big) \,, \\[0.7mm]
\b\b \lambda_{4,5} = P \pm \sqrt{Q} \,,
\eea
where
\bea
P \a=\a \text{\small $\frac 12$} (3 + \epsilon)^2 y \sin^4\! \theta 
+ \text{\small $\frac 12$} (3 + \epsilon)^2 (1 + x) \cos^4\! \theta \nn \\
\a\;\a -  \text{\small $\frac 12$} (3 + \epsilon) (5 + 4 \epsilon) \cos^2\! \theta + (1 + \epsilon) (3 + 2 \epsilon) \,, \\
Q \a=\a \text{\small $\frac 14$} (3 + \epsilon)^4 y^2 \sin^8\! \theta - \text{\small $\frac 12$} (3 + \epsilon)^4 y (1 + x) \sin^4\! \theta \cos^4\! \theta  \nn \\
\a\;\a + \text{\small $\frac 12$} (3 + \epsilon)^3 y\, (7 + 6 \epsilon)  \sin^4\! \theta \cos^2\! \theta - (3 + \epsilon)^2 y\, (1 + \epsilon) (5 + 2 \epsilon) \sin^4\! \theta \nn \\
\a\a + \text{\small $\frac 14$} (3 + \epsilon)^4 (1 + x)^2 \cos^8\! \theta- \text{\small $\frac 12$} (3 + \epsilon)^3  (1+ x) (7 + 6 \epsilon) \cos^6\! \theta \nn \\
\a\;\a + \text{\small $\frac 14$} (3 + \epsilon)^2 \big(69 + 112 \epsilon + 28 \epsilon^2 + 4 x (1 + \epsilon) (5 + 2 \epsilon)\big) \cos^4\! \theta \nn \\[1mm]
\a\;\a - (3 + \epsilon) \big(35 + 79 \epsilon + 44 \epsilon^2 + 8 \epsilon^3\big) \cos^2\! \theta  + (1 + \epsilon)^2 (5 + 2 \epsilon)^2 \,.
\eea
Using the same kind of notation, we also denote
\bea
\b\b m^2_{\rm up} = \lambda_{\rm up}\, m^2_{3/2} \,,\\
\b\b m^2_{\rm low} = \lambda_{\rm low}\, m^2_{3/2} \,.
\eea

In this class of models, it is straightforward to verify all the statements of the previous section concerning the range 
that the mass eigenvalues are allowed to take. For any given value of $\epsilon$, one may make the vacuum as 
stable as possible by first adjusting $\theta$ to the optimal value that allows the upper bound $\lambda_{\rm up}$ 
on the smallest eigenvalue to be maximized, and then adjusting $x$ and $y$ to saturate this value. It turns out that this best 
situation occurs when $\lambda_{1,2} = \lambda_{3} = \lambda_4$, while $\lambda_5$ is always bigger. In table 1 we list
some sample models illustrating this point. We see that in this simple class of models where only two real parameters
are retained among $\alpha_{\hat u \hat v}$ and $\beta$, there are thus always four of the five eigenvalues
that become degenerate when the parameters are adjusted in such a way to saturate the bound defined 
by $\lambda_{\rm up}$. In more general models where additional independent parameters are retained among 
$\alpha_{\hat u \hat v}$ and $\beta$, this feature is expected to disappear.

\begin{table}
\bigskip
\vbox{
$$\vbox{\offinterlineskip
\hrule height 1.1pt
\halign{&\vrule width 1.1pt#
&\strut\quad#\hfil\quad&
\vrule width 1.1pt#
&\strut\quad#\hfil\quad&
\vrule#
&\strut\quad#\hfil\quad&
\vrule#
&\strut\quad#\hfil\quad&
\vrule#
&\strut\quad#\hfil\quad&
\vrule#
&\strut\quad#\hfil\quad&
\vrule#
&\strut\quad#\hfil\quad&
\vrule#
&\strut\quad#\hfil\quad&
\vrule width 1.1pt#\cr
height3pt
&\omit&
&\omit&
&\omit&
&\omit&
&\omit&
&\omit&
&\omit&
&\omit&
\cr
&\hfil $\!\epsilon$&
&\hfil $\theta$&
&\hfil $x$&
&\hfil $y$&
&\hfil $\lambda_{1-4}$&
&\hfil $\lambda_5$&
&\hfil $\lambda_{\rm up}$&
&\hfil $\lambda_{\rm low}$&
\cr
height3pt
&\omit&
&\omit&
&\omit&
&\omit&
&\omit&
&\omit&
&\omit&
&\omit&
\cr
\noalign{\hrule height 1.1pt}
height3pt
&\omit&
&\omit&
&\omit&
&\omit&
&\omit&
&\omit&
&\omit&
&\omit&
\cr
&$\!0$& 
&$\!0.6155\!$& 
&$\!0.0000\!$& 
&$\!0.0000\!$& 
&$\!0.0000\!$& 
&$\!\hspace{6pt}0.000\!$& 
&$\!0.0000\!$& 
&$\!\hspace{6pt}0.000\!$&
\cr
height3pt
&\omit&
&\omit&
&\omit&
&\omit&
&\omit&
&\omit&
&\omit&
&\omit&
\cr
\noalign{\hrule}
height3pt
&\omit&
&\omit&
&\omit&
&\omit&
&\omit&
&\omit&
&\omit&
&\omit&
\cr
&$\!1\!$&
&$\!0.6268\!$&
&$\!0.0799\!$&
&$\!1.8748\!$&
&$\!\hspace{-6pt}\text{--}0.2752\!$&
&$\!\hspace{6pt}7.645\!$&
&$\!\hspace{-6pt}\text{--}0.2752\!$&
&$\!\hspace{6pt}3.210\!$&
\cr 
height3pt 
&\omit&
&\omit& 
&\omit& 
&\omit&
&\omit&
&\omit&
&\omit&
&\omit&
\cr
\noalign{\hrule }
height3pt 
&\omit& 
&\omit&
&\omit& 
&\omit&
&\omit&
&\omit&
&\omit&
&\omit&
\cr
&$\!2\!$& 
&$\!0.6510\!$&
&$\!0.0161\!$& 
&$\!2.5531\!$&
&$\!\hspace{-6pt}\text{--}0.0807\!$&
&$\!19.731\!$& 
&$\!\hspace{-6pt}\text{--}0.0807\!$&
&$\!\hspace{6pt}8.553\!$&
\cr
height3pt 
&\omit&
&\omit& 
&\omit& 
&\omit&
&\omit&
&\omit&
&\omit&
&\omit&
\cr
\noalign{\hrule } 
height3pt 
&\omit&
&\omit& 
&\omit& 
&\omit&
&\omit&
&\omit&
&\omit&
&\omit&
\cr
&$\!3\!$&
&$\!0.6705\!$&
&$\!\hspace{-6pt}\text{--}0.0883\!$&
&$\!2.8275\!$&
&$\!0.5989\!$&
&$\!36.323\!$&
&$\!0.5989\!$&
&$\!16.058\!$&
\cr 
height3pt 
&\omit&
&\omit& 
&\omit& 
&\omit&
&\omit&
&\omit&
&\omit&
&\omit&
\cr
\noalign{\hrule } 
height3pt 
&\omit&
&\omit& 
&\omit& 
&\omit&
&\omit&
&\omit&
&\omit&
&\omit&
\cr
&$\!4$&
&$\!0.6854\!$&
&$\!\hspace{-6pt}\text{--}0.2011\!$&
&$\!2.9547\!$&
&$\!1.7701\!$&
&$\!57.425\!$&
&$\!1.7701\!$&
&$\!25.730\!$&
\cr 
height3pt 
&\omit&
&\omit& 
&\omit& 
&\omit&
&\omit&
&\omit&
&\omit&
&\omit&
\cr
}
\hrule height 1.1pt}
$$
}
\caption{Some explicit examples of models with their spectrum of masses.}
\bigskip
\end{table}

\section{Conclusions}
\setcounter{equation}{0}

In this work we have shown that metastable de Sitter vacua may arise in rather simple N=2 supergravity 
theories with a single hypermultiplet and a single vector multiplet, without Fayet-Iliopoulos terms or 
non-Abelian gauge symmetries, provided that the scalar manifold is suitably curved. A first, crucial requirement 
for this to be possible is that both sectors should be involved in the process of supersymmetry breaking, 
as it is known that in theories with only one hypermultiplet or only one vector multiplet no metastable vacua 
can ever arise. A second, important limitation is that the cosmological constant $V$ should be sufficiently large 
compared to the gravitino mass squared $m^2_{3/2}$ in Planck units. More precisely, for positive $\epsilon = V/m^2_{3/2}$
we showed that metastable de Sitter vacua are only possible if $\epsilon \gsim 2.17$.
As a consequence, the simple de Sitter vacua that we constructed can realize slow-roll inflation with a large 
Hubble parameter corresponding to $\epsilon \gg 1$ but not a late-time vacuum with small cosmological constant 
corresponding to $\epsilon \ll 1$. 

In the simple class of models that we have studied, the quaternionic-K\"ahler manifold describing the hypermultiplet
sector is a generic four-dimensional Przanowski-Tod space possessing at least one triholomorphic isometry. The 
special-K\"ahler manifold describing the vector multiplet sector is a totally general two-dimensional special space 
and the potential is generated by gauging the isometry of the hypermultiplet geometry.
For this class of models, we were able to derive a simple upper bound on the mass of the lightest scalar, which 
depends only on the angle $\theta$ defining the spread of the supersymmetry breaking direction between the two sectors
and the parameter $\epsilon$ related to the cosmological constant. In the particular cases where only the hypermultiplet 
or only the vector multiplet is involved in supersymmetry breaking, corresponding to $\theta \to 0$ and $\theta \to \frac {\pi}2$
respectively, one recovers for any $\epsilon$ the results of \cite{GRLS} and \cite{MANY}, namely the saturable bounds:
\bea
\a\a \min \big\{m^2_i \big\} \le - \text{\small $\frac 13$}\hspace{1pt} m^2_{3/2} - V \;\;\text{(only hyper)} \,, \label{hyp} \\
\a\a \min \big\{m^2_i \big\} \le - 2\hspace{1pt} V \;\;\text{(only vector)} \,. \label{vec}
\eea
In the more general case where both the hyper- and vector multiplets are involved in supersymmetry breaking, 
there exists for any given $\epsilon$ an optimal choice for $\theta$ that maximizes the smallest mass. This allows us
to define a bound depending only on $\epsilon$ for the situation where the two sectors are optimally mixed. 
The exact result for the saturable upper bound on $\min \big\{m^2_i \big\}$ as a function of $\epsilon$ is plotted in figure 2 but 
cannot be expressed in any simple, analytic form. However, we showed that the following simpler bound, which is slightly weaker and 
non-saturable, is also true and gives a good approximation to the exact bound:
\bea
\a\a \min \big\{m^2_i \big\} < -  \text{\small $\frac 12$} \, V +  \text{\small $\frac 14$} \, \frac {V^2}{m^2_{3/2}} \;\;\text{(hyper and vector)} \,.
\label{hypvec}
\eea
Notice that on the right-hand side of this bound, the transition between positive and negative values arises at $\epsilon = 2$,
but as already said the precise critical point for which metastable vacua really become possible is $\epsilon \simeq 2.17$.
It is straightforward to generalize the analysis presented in this paper to study the possibility of getting metastable 
supersymmetry breaking anti de Sitter vacua with a negative cosmological constant $V > - 3\, m^2_{3/2}$ in the same setting. 
In fact, the equations (\ref{m2+-})--({\ref{Y}}) 
defining the bounds $m^2_\pm$ on the scalar masses hold true unchanged also in the case of 
negative $\epsilon$, and one can repeat the same analysis as for positive $\epsilon$. The maximal 
value $m^2_{\rm up}$ that $m^2_-$ can take is found to be positive for 
$\epsilon \in \,\,]\!-\!3,-\frac 32[ \,\, \cup \,\, ]\!-\!1,0[\,$ and negative for $\epsilon \in \,\,]\!-\!\frac 32,1[$, but always 
larger than the Breitenlohner--Freedman bound $\frac 34 V$ for stability \cite{BF}. This implies that metastable 
supersymmetry breaking anti de Sitter vacua may exist for any $\epsilon \in \,\, ]\!-\!3, 0[\,$.

It would be interesting to investigate whether the results found in this paper for models with a single hypermultiplet and 
a single vector multiplet can be extended to theories involving several of these multiplets, at least in the simplest case 
of Abelian gaugings without constant Fayet-Iliopoulos terms. A natural way to try to do this is to look at the mass matrix 
in the subspace of scalar fields defined by the various sGoldstini, along the lines of \cite{GRS1,GRS2,GRS3,GRLS}. 
Indeed, this approach allows one to extend the results (\ref{hyp}) and (\ref{vec}) to theories with an arbitrary number 
of hypermultiplets and an arbitrary number of vector multiplets, respectively. In this way, one might also hope to derive 
a result similar to (\ref{hypvec}) for theories with several hyper- and several vector multiplets. 
A first hint in favor of this comes from the fact that the known examples of metastable de Sitter vacua in such theories \cite{FTV} 
do indeed satisfy this type of bound, albeit in a rather trivial way, thanks to the fact that they all lead to $m_{3/2} = 0$. 
We leave a thorough analysis of this general problem for future work.

\vskip 20pt
\noindent
{\Large \bf Acknowledgements}
\vskip 10pt
\noindent
We would like to thank G.~Dall'Agata and M.~Trigiante for useful comments and discussions.
The research of C.~S. and P.~S. is supported by the Swiss National 
Science Foundation (SNSF) under the grant PP00P2-135164, and that of F.~C.
by the German Science Foundation (DFG) under the Collaborative Research 
Center (SFB) 676 ``Particles, Strings and the Early Universe''.

\end{document}